\documentclass[aps,prd,showpacs,floatfix,preprintnumbers,amsfont,amsmath,amssymb,nofootinbib, superscriptaddress]{revtex4}

\usepackage{graphicx} 
\usepackage[utf8]{inputenc}
\usepackage[english]{babel}
\usepackage{amsmath}
\usepackage{amsfonts}
\usepackage{amssymb}
\usepackage{listings}
\usepackage{lipsum}
\usepackage{mathtools,slashed}
\usepackage{multirow}
\usepackage{datetime}
\usepackage{graphicx}
\usepackage{mathtools}
\usepackage{mathrsfs}
\usepackage{dcolumn}
\usepackage{comment}
\usepackage{multirow}
\usepackage{color} 
\usepackage{lipsum} 

\usepackage{graphicx}
\usepackage{bm}
\usepackage{xcolor}
\usepackage[colorlinks=true,allcolors=blue]{hyperref}

\newcommand{\bea}{\begin{eqnarray}}
\newcommand{\eea}{\end{eqnarray}}

\begin{document}

\title{The universe as a superconductor: How to search for a photon mass}

\author{Andrei Gruzinov}
\email{ag92@nyu.edu}
\affiliation{Center for Cosmology and Particle Physics, Department of Physics, New York University, New York, NY 10003, USA}

\author{Anson Hook}
\email{hook@umd.edu}
\affiliation{Maryland Center for Fundamental Physics, University of Maryland, College Park, MD 20742, USA}

\author{Junwu Huang}
\email{jhuang@perimeterinstitute.ca}
\affiliation{Perimeter Institute for Theoretical Physics, 31 Caroline St. N., Waterloo, Ontario N2L 2Y5, Canada}

\date{\today}

\begin{abstract}
We revisit the bounds on the photon mass when it arises from an Abelian Higgs mechanism rather than a Proca term. Because the higgs is necessarily millicharged and light, the phase of the theory (Meissner, vortex, or symmetry-restored) depends on the ambient magnetic field, so the photon mass an experiment measures depends on the environment it is performed in. We classify existing experiments by the phases in which they apply, in analogy with Type I and Type II superconductors. For Type I photon masses, magnetic fields in galaxies and the intergalactic medium can restore the symmetry everywhere, allowing a vacuum photon mass as large as $0.3 \,{\rm MeV}$. Regions of photon mass as large as $10^{-5}\, {\rm eV}$ could be present in the Universe, and measurement of frequency dependent optical depth can potentially reveal these regions.
For Type II photon masses, we show that the strongest constraints mainly arise from the requirement that the galactic dynamo generates the observed magnetic fields, which we derive by generalizing magnetohydrodynamics to a nonzero photon mass. Since the terrestrial magnetic field restores the symmetry in much of the remaining parameter space, laboratory searches performed inside large magnetically shielded volumes, or in space, could probe photon masses invisible to any experiment on Earth. Measurements of distant light sources, in particular, fast radio bursts, also offer unique sensitivity to parameters invisible to lab experiments.
\end{abstract}

\maketitle

\section{Introduction}

Electromagnetism and gravity are the most important forces that shape our everyday experience. Their force carriers, the photon and the graviton, famously, are the only potentially massless bosons in nature~\cite{ParticleDataGroup:2024cfk,Goldhaber:2008xy}. This important property has been under scrutiny since the time of Coulomb and Cavendish, and so far, there is no experimental evidence that the mass of the photon is non-zero~\cite{Goldhaber:2008xy}. 

Strong limits on the photon mass arise from a variety of experimental searches~\cite{ParticleDataGroup:2024cfk,Goldhaber:2008xy,Bonetti:2017pym,Wang:2023fnn,Romanenko:2023irv,Kalia:2025afc,Williams:1971ms,Ryutov:2007zz,Chibisov:1976mm,10.1143/PTPS.11.1,Lakes:1998mi,Luo:2003rz,Dvali:2025mqr}, which can be classified into the following four types. The first class of constraints relies on photon propagation, testing for modifications to the dispersion relation~\cite{Bonetti:2017pym,Wang:2023fnn}. It also encompasses searches for the longitudinal mode of a massive photon, which is weakly coupled and can therefore propagate through walls~\cite{Romanenko:2023irv,Kalia:2025afc}. The second class consists of laboratory tests of Coulomb's law, i.e. the non-observation of an exponential fall-off of electric fields with distance from their source~\cite{Williams:1971ms}. A third class of constraints come from magnetic fields, whose dynamics and large-scale structure are altered substantially by a nonzero photon mass~\cite{Ryutov:2007zz}. A fourth class relies on a large vector potential $A^{\mu}$, which includes both energy-density arguments, that is, the Chibosov-Yamaguchi bound (CY)~\cite{Chibisov:1976mm,10.1143/PTPS.11.1}, as well as torques proportional to the vector potential (Lakes tests)~\cite{Lakes:1998mi,Luo:2003rz}. 

While there is no evidence for a photon mass in vacuum, in the presence of matter, it does acquire a non-zero mass. In a superconductor, $U(1)_{\rm EM}$ is Higgsed, leading to the screening of the electric field and the Meissner effect~\cite{tinkham2004introduction}. Whereas a substantial fraction of materials become a superconductor at zero temperature, we have yet to find a room temperature superconductor because superconductivity can be easily destroyed by environmental effects, such as non-zero temperatures and magnetic fields. This subtlety was first appreciated in the context of the Standard Model photon in Ref.~\cite{Adelberger:2003qx}, where it was pointed out that if the photon mass arises from the Abelian Higgs model, with action~\footnote{We adopt the convention of~\cite{tinkham2004introduction,East:2022rsi} to better match the critical field definitions, instead of a canonically normalized kinetic term of the complex field $\Phi$. This distinction is invisible on the scales of our figures and do not affect any conclusions.}
\begin{equation}
    \mathcal{L} = \frac{1}{2}\left|D_{\mu} \Phi\right|^2 -\frac{1}{4}F^{\mu\nu}F_{\mu\nu}-\frac{\lambda}{4}\left(|\Phi|^2-v^2\right)^2 + e A_{\mu} J^{\mu}_{\rm EM}, 
\end{equation}
then the strongest constraint on the Proca photon mass, the Chibisov-Yamaguchi bound~\cite{Chibisov:1976mm,10.1143/PTPS.11.1}, can be circumvented as long as the higgs mode is light enough (see also the photon mass page in~\cite{ParticleDataGroup:2024cfk}). Here, $D_{\mu} = \partial_{\mu} - i g A_{\mu}$, 
$J^{\mu}_{\rm EM}$ contains all the Standard Model charged particles, and $g = Q_h e$ with $Q_h \ll 1$, since the higgs field $\Phi = \rho e^{i \theta}$ is necessarily millicharged under the $U(1)_{\rm EM}$. In vacuum, $\langle \rho\rangle = v $, the photon mass is $ m_A = g v$ and the higgs mass $m_h=\sqrt{2\lambda} v$.

Unlike the case of a Proca mass, which corresponds to the limit of $m_h \rightarrow \infty, m_A = {\rm const}$, including the Abelian Higgs mechanism leads to several distinct behaviors. The most important is that as a function of the strength of the magnetic field and current, the system can be in a few different phases. The classification, similar to the case of a superconductor, depends on the ratio $\lambda/g^2 = m_h^2/2m_A^2$. 
When $\lambda/g^2 < 1/2$, similar to a Type I superconductor, there is one critical field $B_c = \lambda v^2/g$, above which $U(1)_{\rm EM}$ symmetry is restored. On the other hand, 
when $\lambda/g^2 > 1/2$, similar to a Type II superconductor, there are three distinct critical fields
\begin{itemize}
    \item $B_{\rm c1} \simeq g v^2$: Magnetic field strength of a vortex, above which strings are energetically preferred.
\item $B_{\rm sh} = \lambda^{1/2} v^2$: Critical field strength for vortex formation to occur classically~\cite{Galaiko1966FormationOV,East:2022rsi}.
    \item $B_{\rm c2} = \lambda v^2/g$: Critical field strength for symmetry restoration.
\end{itemize}
For both Type I and Type II, there is an equally important quantity, the critical (London) current for formation of Pearl (global) vortices~\cite{1964ApPhL...5...65P,Fedderke:2025sic}, $J_c = m_A m_h v$, which can subsequently lead to the formation of Abrikosov vortices, as well as full symmetry restoration depending on the details of the system. 

In the following, we will borrow the terminology from the superconductor literature and refer to the case of $\lambda/g^2 < 1/2$ as a Type I photon mass and $\lambda/g^2 > 1/2$ as a Type II photon mass.
For Type I and Type II photon masses, depending on the magnetic field strength,  different systems can be in different phases, drastically affecting our ability to search for a non-zero photon mass. 
The Proca Lagrangian applies in the limit of zero magnetic field, or more precisely, in the Meissner phase when the magnetic field strength is less than $B_c$ ($B_{\rm c1}$) for $\lambda/g^2  < 1/2$ ($\lambda/g^2 > 1/2$). 
Meanwhile, if the magnetic field strength is such that the system resides in the vortex phase ($B_{\rm sh} < B <B_{\rm c2}$ and $\lambda/g^2 > 1/2$)~\footnote{Throughout the paper, we write $B \equiv |\vec B|$, and similarly for vector potentials and current densities.}, then the searches which are based on energy density arguments and magnetic field dynamics or structure generally do not apply.
If the magnetic field strength is even larger than $B_c = B_{\rm c2}$, then the photon would behave like a perfect massless particle in this system, and no photon mass bound would apply. 

In this paper, we present a systematic study of Type I and Type II photon mass models as a function of their three parameters, the photon mass $m_A$, the quartic coupling $\lambda$ and the electric charge $g$ of the higgs field. The goal is to display clearly the regions of parameter space that are open to being explored, to identify which type of experiments can probe the remaining parameter space, and to motivate new searches that can fill the gap. A most striking conclusion from this extension is that a (vacuum) photon mass can be as large as $3\times 10^5\,{\rm eV}$.
In Sec.~\ref{sec:mag}, we provide a phase diagram for the theory.  This illustrates which experiments can effectively search for a photon mass in the various regions of parameter space.
In Sec.~\ref{sec:plasma}, we discuss how the galactic dynamo changes due to a Type II photon mass and how the galactic magnetic field can still be used to place a constraint even in the vortex phase. Finally, and most importantly, in Sec.~\ref{sec:parameter} we discuss new experimental methods by which one can explore these regions, highlighting the importance of performing laboratory photon mass measurements in space, as well as cosmological observations that exploit the correlation between a photon mass and cosmic void. We conclude in Sec.~\ref{sec:con}.

\section{The Phase Diagram for a Massive Photon}\label{sec:mag}

The central feature of the Abelian Higgs photon mass, and what distinguishes it from a pure Proca mass, is that the response of the system to an external magnetic field depends on the field strength. Experimentally, this implies that the photon mass an experiment measures depends on the environment it resides in. Before we delve into the individual experiments, we first outline what types of measurements are possible in which phase. 

\paragraph{Meissner Phase}
In this phase, the Type I \& II photon mass behaves exactly like the Proca action and all of the constraints derived based on the Proca action apply.

\paragraph{Abrikosov Vortex Phase}
In this phase, vortex formation discharges the vector potential $A$, preventing it from growing as large as it otherwise would have.  The result is that experiments that target a large $A_{\mu}$ through either a large current~\cite{Lakes:1998mi,Luo:2003rz} or energy density~\cite{Chibisov:1976mm,10.1143/PTPS.11.1,Ryutov:2007zz} are rendered insensitive.

\paragraph{Symmetry Restored Phase} In the symmetry restored phase, all measurements of the photon mass give zero and the vacuum photon mass cannot be measured. The only trace of this mechanism hides in the very light millicharged particles. 

In the following, we describe the phase diagrams of Type I and Type II cases separately.

\subsection{Type I}\label{sec:2typeI}

\begin{figure}
    \centering
    \includegraphics[width=0.5\linewidth]{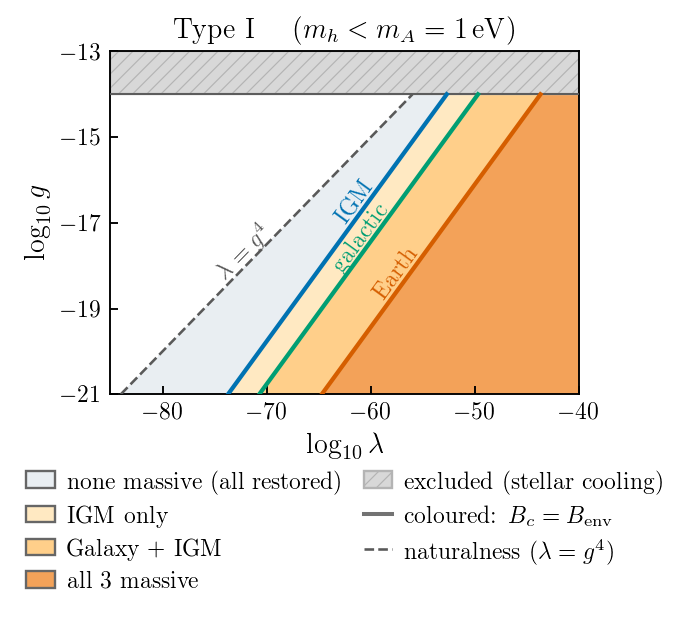}
    \caption{Type I photon mass phase diagram for $m_A = 1\,{\rm eV}$.  Shaded regions indicate where photon is massive or massless in different environments.  The region to the left of the dashed $\lambda = g^4$ line is not theoretically accessible without extreme functional tuning. To the left of the blue line, a nanoGauss magnetic field (as what might be present in the IGM) would be able to restore the symmetry.  To the left of the green line, the Milky Way's observed magnetic field restores the symmetry~\cite{2026arXiv260602149H}.  To the left of the red line, the Earth's observed magnetic field restores the symmetry. A maximal higgs mass of $m_h = 3\times 10^{-13} \,{\rm eV}$ is reached at the intersection between the blue line and the boundary of the stellar cooling constraint }
    \label{fig:type1}
\end{figure}

For a Type I photon mass, there are only two options.  Either the photon mass in the environment is non-zero and all bounds apply, or the symmetry is restored and all bounds do not apply.  The phase diagram for a Type I photon mass is shown in Fig.~\ref{fig:type1}.  To the right of the red Earth (green galactic) line, all Earth (galactic) based  constraints apply.  Meanwhile to its left, Earth (galactic) based constraints no longer apply due to symmetry restoration.  While the value of the intergalactic magnetic field is not known, under reasonable assumptions, it can be as large as nano-Gauss (nG).  To the left of the blue IGM line, even a small nG magnetic field would restore the $U(1)$ symmetry and render all constraints invalid.

Strikingly, it is actually possible for the photon to have a mass in vacuum that is as large as $3\times 10^5 \,{\rm eV}$. In the limit of $\lambda \rightarrow 0$, we have $B_c \rightarrow 0$, which shows that, due to the extreme flatness of the Higgs potential, any small magnetic field can in principle restore the symmetry.

Fully realizing this interesting region is hampered by the fact that it is not possible to take $\lambda \rightarrow 0$ at all energy scales without extreme fine-tuning. 
Due to 1-loop renormalization running, $\lambda$ changes at the $\mathcal{O}(g^4)$ level.  To set $\lambda \ll g^4$ at all energies would therefore require a separate cancellation at every scale, not just one. This is qualitatively worse than the well-known fine-tuning of the Higgs mass,
where the correction is large but is a single number that
can be cancelled once at the cutoff. This systematic tuning would need to be done for an infinite tower of terms at all energies and temperatures, which we will not entertain in this paper. Instead, saturating the requirement of $\lambda/g^4 \geq 1$, and $B_{\rm IGM} \geq B_c$ allows us to get a maximal photon mass of
\begin{equation}
    m_A \approx \left(v B_{\rm IGM} \right)^{1/3} \leq \left(M_{\rm pl} B_{\rm IGM} \right)^{1/3} = 3\times 10^5 \,{\rm eV} \left(\frac{ B_{\rm IGM}}{{\rm nG}}\right)^{1/3},
\end{equation}
if the vev of the $U(1)_{\rm EM}$ higgs saturates the Planck scale $M_{\rm pl}$, and only the higgs mass parameter is tuned, similar to the situation of the Standard Model Higgs. In the UV theory of $SU(2)_L \times U(1)_Y$, this higgs is millicharged under $U(1)_Y$, and the vev should be interpreted as Higgsing $U(1)_Y$ extremely weakly. Parameters that saturate this limit are $g\sim 10^{-22}$ and $\lambda \sim 10^{-88}$, and the millicharged higgs mass is about $3 \times 10^{-17}\,{\rm eV}$. 

Another important feature to notice is that there is a maximal $m_h$, corresponding to saturating both the millicharged particle limit of $g \simeq 10^{-14}$ and the magnetic field in extra galactic space $B_{\rm IGM} = B_c$. At this point, we have $m_h \sim g^{1/2} B_{\rm IGM}^{1/2} \approx 3\times 10^{-13} \,{\rm eV}$. This maximal higgs mass is independent of the photon mass of interest in the range of $m_A \lesssim 10 \,{\rm eV}$.

A final feature of a Type I photon mass, is that strong constraints from the galactic magnetic field can only be avoided if $B_{g} > B_c$, where the symmetry is restored in the galaxy. This combined with $m_h < m_A$, unfortunately also implies that symmetry is always restored on Earth, since shielding a region of $\gtrsim (10^3 \,{\rm km})^3$ seems unlikely.

\subsection{Type II}\label{sec:2typeII}

In this subsection, we focus on the Type II parameter space as a function of magnetic field. 
The parameter space is shaped by the following conditions. Firstly, a Type II mass by definition has $\lambda/g^2>1/2$.  Secondly, $g \lesssim 10^{-14}$ since the higgs is a millicharged particle, and light millicharged particles are constrained by stellar cooling~\cite{Fung:2023euv}. Lastly, depending a bit on the plasma effects discussed in Sec.~\ref{sec:plasma}, the strength of the magnetic field of various environment as compared to the various critical fields ($B_{\rm sh}$ and $B_{\rm c2}$) can also change whether various constraints apply. In particular, whether the Chibisov-Yamaguchi bound~\cite{10.1143/PTPS.11.1,Chibisov:1976mm} applies or not depends on how the galactic magnetic field compares against the super-heating field strength ($B_{\rm sh}$)\footnote{Here we keep the requirement of $B> B_{\rm sh}$.  This bound can strengthen or weaken depending on plasma dynamics.  For example, it can strengthen due to plasma dynamics we will elaborate on in Sec.~\ref{sec:plasma}.  It can also weaken if we assume pre-existing vortices produced by early universe dynamics or if vortices were produced locally at strong B-field regions, for example stars, supernova and black holes, and were stretched into the form it has in the galaxy today. 
To simplify matters, hereafter we assume that the galactic magnetic fields are produced through the dynamo process, and vortices form in-situ. A small number of vortices could come from early universe dynamics that leave $\mathcal{O}(1)$ strings per Hubble patch (scaling solution)~\cite{KIBBLE1980183}, however, it is inconceivable that these strings can be responsible for turning all the galaxies in the Universe into the vortex phase.}. This situation was incorrectly described in~\cite{Adelberger:2003qx} for non-trivial reasons, see App.~\ref{Sec:inside} for more details.

\begin{figure}
    \centering
    \includegraphics[width=0.99\linewidth]{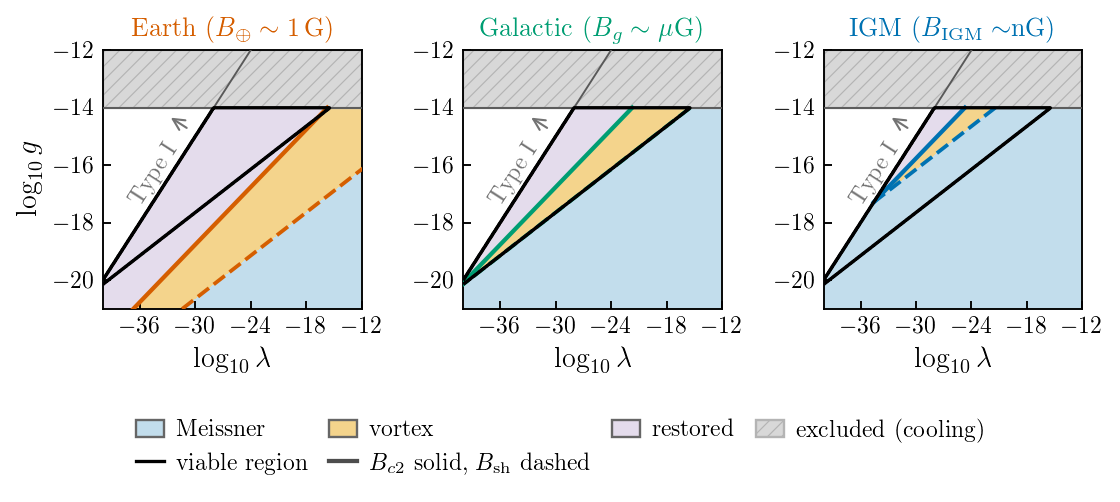}
    \caption{We show the phase space diagram for a Type II photon mass ($m_A = 10^{-14} \,$ eV) for three separate environments: Earth (red), the Milky Way (green), and the intergalactic medium (blue).  The region inside of the black triangle is the allowed (non-excluded) parameter space.  The white region is the Type I parameter region and is described in Fig.~\ref{fig:type1}.
    To the left of the solid line (shaded red region), the system is in the symmetry restored phase.  
    To the left of the dashed line (shaded orange region), the system is in the vortex phase.
    To the right of the dashed line (shaded blue region), the system behaves as if the mass were a Proca mass term.  }
    \label{fig:typeIIenv}
\end{figure}

For example, for a fixed $m_A = 10^{-14} \,$ eV, the phenomenologically viable parameter space is shown in Fig~\ref{fig:typeIIenv}.  The region enclosed by the thick black lines ($\lambda/g^2>1/2$, $g< 10^{-14}$, and $B_{\rm galaxy} > B_{\rm sh}$) is the available parameter space.  The shaded red region is the symmetry restored phase and no photon mass bounds apply.  The shaded orange region is the vortex phase, where only a limited set of constraints apply.  Finally, the blue shaded region is the Meissner phase where all constraints apply.  Interestingly, in all of the allowed parameter space for this photon mass, Earth based constraints, no matter how sensitive, would never be able to measure a non-zero photon mass as the symmetry is restored on Earth.  In contrast, the galactic magnetic field is not sufficient to fully restore the symmetry, and searches based on photon propagation and Coulomb's tests can in principle measure a non-zero photon mass of this size.
This observation will lead us to propose a new class of experiments in Sec.~\ref{sec:parameter} designed to probe this parameter space.

\begin{figure}
    \centering
    \includegraphics[width=0.99\linewidth]{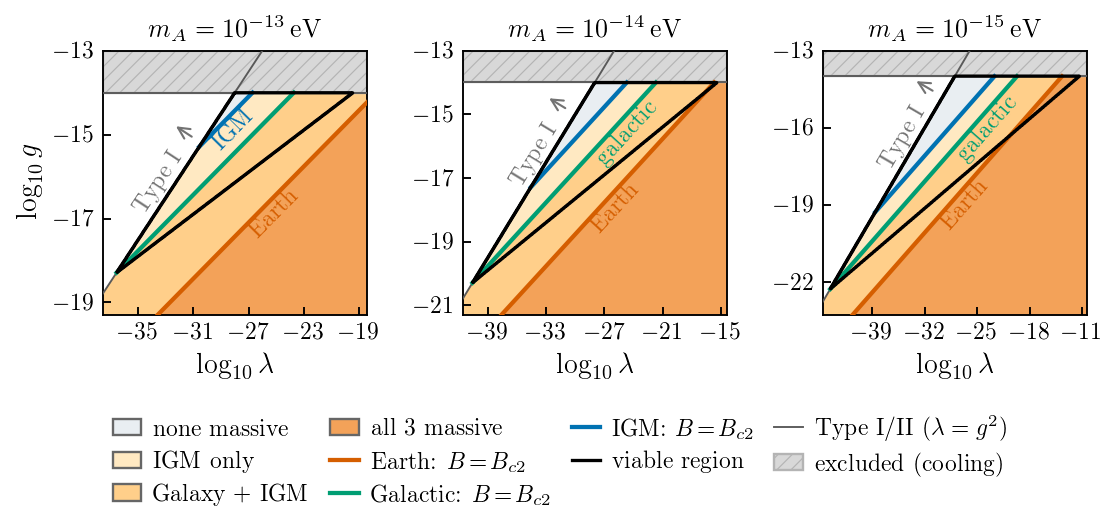}
    \caption{The symmetry restoration phase space diagram for three values of a Type II photon mass.  
    To the left of the solid blue line, the IGM may be in the symmetry restored phase.  To the left of the solid green/red lines, the Milky Way/Earth would be in the symmetry restored phase.  As before, the allowed parameter region is inside the black triangle.  }
    \label{fig:typeIIsweep}
\end{figure}

Fig.~\ref{fig:typeIIsweep} shows the parameter region from a different view point.  The figure focuses on symmetry restoration as it is what is important for Earth based searches. For a few values of a Type II photon mass, we show the allowed parameter space (the black triangle).  To the left of the blue/green/red lines, indicates when symmetry is restored by the magnetic fields of the IGM/Milky Way/Earth.  As one increases the photon mass, the available parameter space shrinks.
For $m_A \gtrsim 10^{-11} {\,\rm eV}$, the whole parameter space is closed. For lighter $m_A$, parameter space opens up with increasing $\lambda$ and $\lambda/g^2$. For $10^{-14} {\,\rm eV} \lesssim m_A \lesssim 10^{-11} {\,\rm eV}$, Earth's magnetic field (taken to be 1 Gauss) is strong enough that in all of the remaining parameter space, it is stronger than even the critical field $B_{\rm c2}$ (indicated by the thick red line), and the Earth would be in a symmetry restored phase, instead of a vortex phase, meaning that the photon mass would be zero on the Earth, and no experiment on Earth can be sensitive to this massive photon without shielding the magnetic field. As the photon mass decreases, the red curve separates the triangle into two parts. In the part to the left of the red line, the photon would be massless on the Earth, while to the right of the red line, the Earth would be in a vortex phase and experiments on Earth can probe viable parameter space.

The surprising result of this simple analysis is that, although one would expect that a photon mass of $10^{-14} {\,\rm eV} $ would violate many lab based constraints, in this model, such a large photon mass is allowed because the photon is actually massless on the Earth due to symmetry restoration, while remaining massive away from the Earth.
In fact, many of the lab probes are performed in environments with large field amplitudes. This design can by itself restore the symmetry and destroy the signal they are attempting to measure.
From a different perspective, this suggests we might want to consider performing a Coulomb test on future space based missions, such as Voyager, where the magnetic field is weaker. In contrast to conventional wisdom, cosmology and FRBs~\cite{Bonetti:2017pym,Wang:2023fnn} may actually provide a uniquely robust probe of the parameter space~\cite{ParticleDataGroup:2024cfk}.

Whereas the Earth magnetic field can be strong enough to create a region of symmetry restoration even for  $m_A = 10^{-11} {\,\rm eV}$, a weaker magnetic field in the inter galactic medium can only fully restore the symmetry for a much lower $m_A$. 
As shown in Fig~\ref{fig:typeIIsweep}, for $m_A =  10^{-13} {\,\rm eV}$ and $g< 10^{-14}$, there is almost no parameter space where the symmetry can be fully restored in an extra galactic field of ${\,\rm nG}$. Therefore, given the millicharged particle constraint, the symmetry cannot be restored everywhere in the universe for $m_A \geq  3\times10^{-13} {\,\rm eV}$. In other words, in all of the parameter space below the $g< 10^{-14}$ and $\lambda \geq g^2/2$, the inter-galactic medium wants to stay in the Meissner or the Vortex lattice phase. Given that the constraint from FRB of $m_A \gtrsim 10^{-14}  {\,\rm eV}$ would apply if the symmetry is not fully restored in extra galactic space, even in the Abelian Higgs model, there is an upper limit on the photon mass that is $m_A \lesssim  3\times 10^{-13} {\,\rm eV}$ in the Type II case. There is also the possibility that the photon is massive only outside of galaxies, and one needs a new analysis pipeline for this case. More details will be provided in Sec.~\ref{sec:parameter}

\section{The Photon mass and a magnetized plasma}\label{sec:plasma}

In the previous section, we discussed the behavior of the Abelian Higgs sector in vacuum. In this section, we turn to its behavior in a plasma background, focusing on the Type II parameter space, where a mechanism to produce magnetic field starting from the symmetry-broken phase is necessary~\footnote{In the Type I parameter space of the most interest (large photon mass), the universe cannot ever be in the broken phase throughout the early universe evolution. However, due to the extreme lightness of the higgs, the thermal correction from the CMB photon bath is in fact sufficient to ensure this, and the magnetic field generation in that case proceeds as in the ordinary case (see Fig.~\ref{Fig:Type1} and Fig.~\ref{Fig:Type2}).}. To avoid excessive review in the main text, we defer a discussion of the basic behavior of magnetostatics with $m_A \ne 0$ to App.~\ref{Sec:static}.
Here, we just summarize two main qualitative points. Firstly, if the source of the magnetic field lies outside the region with a massive photon, then $B \simeq  m_A A$, and hence $B \simeq J_L/m_A$, is a generic expectation for the relation between the magnetic field and the London current $J_L$ at the same location. In this case, as in condensed matter systems, the superheating field $B_{\rm sh} = \lambda^{1/2} v^2$ and the critical London current $J_c \simeq m_A B_{\rm sh}$ are generally reached simultaneously, and the conditions in Sec.~\ref{sec:mag} apply. Secondly, if the current density that sources the magnetic field lies inside the region with $m_A\neq 0$, which is closer to the situation inside a galaxy, then the Meissner effect dictates that the critical London current is generally reached before the superheating field $B_{\rm sh}$. In this case, it is essential to understand galactic magnetohydrodynamics and how it is modified in the presence of a photon mass.

In this section, we first present how the equations describing the generation of magnetic fields are modified by a nonzero photon mass. We then discuss how the critical conditions for vortex formation change relative to Sec.~\ref{sec:mag}, and how the magnetic field grows once vortices have formed.

\subsection{Induction equation with a photon mass}

We begin with the dynamo.\footnote{A review of the relevant $m_A = 0$ magnetohydrodynamics is given in App.~\ref{Sec:review}.}
If the photon is massive, in the Meissner phase, the non-relativistic Amp\`ere's law is modified by a Proca term to
\begin{equation} \label{Eq:ampere_m}
\nabla\times\vec B = \vec j_{\rm{free}} - m_A^2\vec A \, .
\end{equation}
Since standard model charged particles only respond to electric and magnetic fields, Ohm's law is unchanged.  Using the modified Ampere's law, the Bianchi identity, and Ohm's law gives the governing equation
\begin{equation} \label{Eq:new_master}
    \frac{\partial\vec B}{\partial t} = \nabla\times(\vec u\times\vec B) + \eta\, \left ( \nabla^2 - m_A^2 \right ) \vec B,
\end{equation}
where $\vec u$ is the fluid velocity. Similar to the $m_A =0$ case, the first term on the right can be rewritten as $(\vec B\cdot \nabla) \vec u -(\vec u\cdot \nabla)\vec B - \vec B (\nabla\cdot\vec u )$, where the first of the three terms is the dynamo term, and determines the rate of the growth of the magnetic field.  
The second term in Eq.~\ref{Eq:new_master}, proportional to the
magnetic diffusivity $\eta = 1/\sigma$, damps the growth of the
magnetic field.

The effect of the photon mass is an additional contribution to the damping of the magnetic field. 
In absence of the dynamo term, the magnetic field decays exponentially
\begin{equation}
\frac{\partial\vec B}{\partial t} = \eta\,(\nabla^2-m_A^2) \vec B, \qquad B(t) = e^{- \eta (k^2 + m_A^2) t} e^{i k x} B (t=0)\, .
\end{equation}
The rate of decay is enhanced compared to the $m_A=0$ case because the external currents need to be larger to generate the same magnetic field as when $m_A=0$ due to the Meissner effect, and consequently, an enhanced Ohmic diffusion rate.

Since the dynamo term is unmodified, magnetic field growth can be much
harder to achieve when $m_A \neq 0$.
Motivated by this, we define a new Magnetic Reynolds Number by comparing the dynamo term and the damping term in Eq.~\ref{Eq:new_master} 
\bea
D_m \equiv \frac{ u/L}{\eta(1/L^2+m_A^2)},
\eea
which depends on a characteristic length scale $L$. Throughout the rest of the discussion
we are mainly interested in the regime $m_A L \geq 1$, in which
\begin{equation}
    D_m \lesssim \frac{u}{L m_A^2 \eta} = \frac{R_m}{m_A^2 L^2} \, ,
\end{equation}
where $R_m = uL/\eta$ is the ordinary Magnetic Reynolds Number. Evidently, $D_m \ll R_m$ when $m_A L \gg 1$. 

It is important to emphasize that Eq.~\ref{Eq:ampere_m}, and therefore also Eq.~\ref{Eq:new_master}, only applies in the Meissner phase.  Any phase that contains higgs dependent dynamics, such as the vortex or symmetry restored phases, invalidates the Proca EFT and its associated equations of motion.  Once vortices form, and/or symmetry is restored, the induction equation actually returns to its original form. This statement is trivial in the symmetry restored phase since the photon becomes massless, and the London current vanishes. In the vortex phase, the vector potential $A$ is bounded from above by $\sim B/m_A$, and the London current cannot grow beyond $J_c$. The physical reason is that the vortices {\it discharge} the vector potential~\cite{Fedderke:2025sic,Tingyu,tinkham2004introduction}. Mathematically, the vector potential can be written explicitly as a combination of the transverse $A^T_\mu$ and the longitudinal mode $\partial_\mu \theta$ as $A_\mu = A^T_\mu-\partial_\mu \theta/g$. At the core of a vortex $\nabla \times\nabla \theta \propto \delta^2(r)$ and these delta functions sum to cancel the effect of the London current on scales much larger than the vortex separation~\cite{tinkham2004introduction}.

This cancellation can be better understood from the perspective of the vortex phase. At the coarse-grained level, averaged over scales $L$ much larger than the vortex separation, the system consists of bundles of vortices and the magnetic field around them, resembling the Abrikosov lattice. Taking the curl of the London current in Eq.~\ref{Eq:ampere_m} and integrating over the cross section area $a$ of a single vortex line gives
\begin{eqnarray}\label{eq:cancellation}
\int \langle \vec \nabla \times \vec J_L \rangle \cdot {\rm d} \vec a &=& m_A^2 \int \vec \nabla \times \vec A \cdot {\rm d}\vec a = m_A^2 \int \vec \nabla \times \vec A_T \cdot {\rm d}\vec a - \frac{m_A^2}{g} \int \vec \nabla \times \vec \nabla \theta \cdot {\rm d} \vec a \nonumber \\
&=& m_A^2 \int \vec B \cdot {\rm d} \vec a - m_A^2 \int \frac{2 \pi}{g} \delta^2(r) {\rm d}a = 0,
\end{eqnarray}
showing that the London current averages to zero on large scales. In this phase, the vortex cores carry a small fraction $\mathcal{O}(B_{\rm c1}/B\leq g/\lambda^{1/2})$ of the total magnetic field energy density, and do not affect the magnetic field dynamics on large scales. This is the reason behind the invalidation of the Chibisov-Yamaguchi bound~\cite{Adelberger:2003qx}, as well as the vanishing of the London current in the vortex phase.

\subsection{Magnetic field production in the galaxy}

The new magnetic Reynolds number is important as the small scale dynamo (the Stretch-Twist-Fold mechanism, which operates on length scales where the fluid is turbulent), requires flux freezing.  The criteria for flux freezing is now $D_m \gg 1$ instead of $R_m \gg 1$.  Unlike $R_m$, which grows with $L$, $D_m$ shrinks at large $L > 1/m
_A$, resulting in a limited range of $L$ for which the dynamo can occur.
The bound on the photon mass, as a result, comes from the requirement that the dynamo process can generate the observed galactic magnetic fields of sufficient strength $B \sim \,{\rm \mu G}$ on sufficiently large scales $\ell_0 \sim  \,{\rm kpc}$. 

The origin of the galactic field is debatable. Here we assume the
``standard scenario": supernova-driven turbulence amplifies the
magnetic field to rough equipartition on all scales $\lesssim$ 1 kpc,
independent of the seed field, see e.g. Ref.~\cite{2017MNRAS.469.3185P}. It is often claimed that
a primordial magnetic field is absolutely necessary as a seed for the
galactic dynamo. If it were so, we would also have to discuss how the
primordial field is affected by the photon mass. However, as proposed
in Ref.~\cite{Gruzinov_2001}, the primordial seed is not a must. 

For a massless photon,  in the standard scenario, the actual seed for
the galactic dynamo (primordial magnetic field, protogalactic
shocks, magnetized stellar winds, supernova shocks, or something else) is unknown, but should be irrelevant; different seeds merely give
different routes to the same ultimate equipartition. For the type II
massive case, however, the actual seed for the galactic dynamo may be
relevant, as different seeds may lead to different Higgs fields, and to
different final outcomes.  The following discussion is therefore
highly uncertain, as it does NOT specify the seed. Dynamo with a
higgsed photon is a difficult and potentially interesting unsolved
problem.

There are two paths for the galaxy to generate such magnetic field. The first is to satisfy the condition\footnote{The threshold for magnetic field growth is likely not strictly at $D_m = 1$. Our numerical studies verified that the magnetic field grows at least when  $D_m \geq 10$. Given the $\mathcal{O}(1)$ uncertainties in various other quantities that describe the galactic plasma, we adopt $D_m = 1$ for simplicity.}
\begin{equation} \label{Eq:Dm constraint}
    D_m(L = \ell_0) \gtrsim 1 \,
\end{equation}
directly, which requires that 
\begin{equation}\label{eq:path1}
    m_A \leq \sqrt{\frac{u_0}{\ell_0\eta}}\approx 6 \times 10^{-16} \,{\rm eV}\left(\frac{u_0}{10^{-3}}\right)^{1/2}\left(\frac{{\rm kpc}}{\ell_0}\right)^{1/2}
\left(\frac{\eta_{\rm SP}}{\eta}\right)^{1/2},
\end{equation}
where $\eta_{\rm SP} = \frac{\sqrt{2\pi}}{3\pi\gamma_E}\frac{\alpha_{\rm EM}m_e^{1/2}\ln\Lambda}{T_e^{3/2}}$ is the Spitzer conductivity, $\gamma_E\approx 1.96$, $\ln\Lambda \approx 20$ the Coulomb logarithm, and $T_e$ is chosen to be  $10^4 \,{\rm K}$~\cite{BoydSanderson}. 
When Eq.~\ref{eq:path1} is satisfied, the dynamo process can happen similarly to the $m_A= 0$ case on the largest scales until London current grows to the critical current $J_c$, and vortices form in the galaxy. Note that even in this case, the condition is already significantly weaker than the Chibisov-Yamaguchi bound.

There is a second, two-stage path to the production of large-scale magnetic fields. In the first stage, the field grows rapidly on small scales until vortices form. Because the growth rate is largest on small scales, this occurs regardless of whether the photon mass vanishes. In the second stage, the magnetic field on large scale proceed to grow in the vortex phase, where the additional contribution to diffusion damping already switched off (see Eq.~\ref{eq:cancellation} and the surrounding discussion).

This suggests that as long as there exists a length scale at which the dynamo can occur until the resulting magnetic field is large enough to produce strings, then the galactic magnetic field can be produced. The smallest scale where fluid turbulence cascades down to is the diffusion scale, derived to be $\ell_\nu \simeq 0.01 {\rm pc}$ (see derivation in App.~\ref{Sec:review}). As reviewed in App.~\ref{Sec:review}, at this scale, the velocity $u_\nu  = u_0 (\ell_\nu/\ell_0)^{1/3} $ following the Kolmogorov scaling, and $D_m (L = \ell_\nu) > 1$ is satisfied for~\cite{landau1987fluid} 
\begin{equation}\label{eq:path2}
    m_A \leq 3\times 10^{-14} \,{\rm eV}
\end{equation}
for the same parameters as in Eq.~\ref{eq:path1}.

While Eq.~\ref{eq:path2} ensures the initial growth of the magnetic field in the galaxy, this magnetic field only grows until equipartition on the corresponding length scale, i.~e., $\beta \rho_{\rm EM} \simeq \rho u_{\nu}^2$ for a plasma beta $\beta \sim 10 - 100$, and $\rho \simeq \, {\rm GeV/cm^3}$ the baryon density in the ISM. Since most of the electromagnetic energy in the Meissner phase is stored in $m_A^2 A_{\mu} A^{\mu}$, and that $B_g^2 \approx \rho u_0^2 /\beta $ in the galaxy, we can find the current density at the diffusion scale when equipartition is reached to be
\begin{equation}\label{eq:j}
    j \approx m_A B_g (\ell_{\nu}/\ell_{0})^{1/3}.
\end{equation}
This current density needs to be larger than the critical current $J_c \simeq m_A B_{\rm sh}$ for strings to be produced in this medium. This led us to the requirement that
\begin{equation}
    B_g > B_{\rm sh} (\ell_{\nu}/\ell_{0})^{-1/3} \approx 46 B_{\rm sh} \left(\frac{\ell_{0}/\ell_{\nu}}{10^5}\right)^{1/3} \, ,
\end{equation}
mildly stronger than the limit derived in the vacuum case. For $ 6 \times 10^{-16} \,{\rm eV} \leq m_A \leq 3\times 10^{-14} \,{\rm eV}$, the largest magnetic field and London current is reached at a scale in between the diffusion scale and an intermediate length scale, and hence vortices will form at this intermediate scale. In this range of $m_A$, the two paths to vortex formation can be summarized into a single condition
\begin{equation}
 J_c \simeq m_A B_{\rm sh} \leq \sqrt{u_0/\ell_0 \eta} B_{g} \approx 10^{-9}{\rm /cm^3} \, .
\end{equation}
This constraint can be satisfied in the interstellar medium, which has number densities of order ${\rm 1/cm^3}$.  This constraint is shown in Fig.~\ref{Fig:Type2}.

Before we close this section, we acknowledge that we have adopted the Spitzer conductivity as the main source of magnetic field dissipation, which relies on the assumption that Coulomb collision, instead of wave-wave interaction, sets the mean free path, and dissipation rate of charged particles. This might be a reasonable assumption for the evolution leading up to vortex formation because unlike the ordinary dynamo process, the magnetic field actually does not grow significantly as energy density is transferred from the turbulent fluid to the electromagnetic sector. Rather, most of the energy density goes to the London current, which does not directly backreact on the charged particles. A detailed numerical study would be helpful for verifying this expectation.

\section{Parameter space and experimental prospect}\label{sec:parameter}

In this section, we first discuss the current constraints on Type I \& II photon masses, including the limits from lab experiments, as well as the updated limits from galactic magnetic field in the previous section. We then discuss new experiments and analysis that one could do to explore new regions of parameter space.

\subsection{Current Constraints}

\subsubsection{Type I photon mass ($\lambda/g^2 <1/2$)}

\begin{figure}
    \centering
    \includegraphics[width=0.80\linewidth]{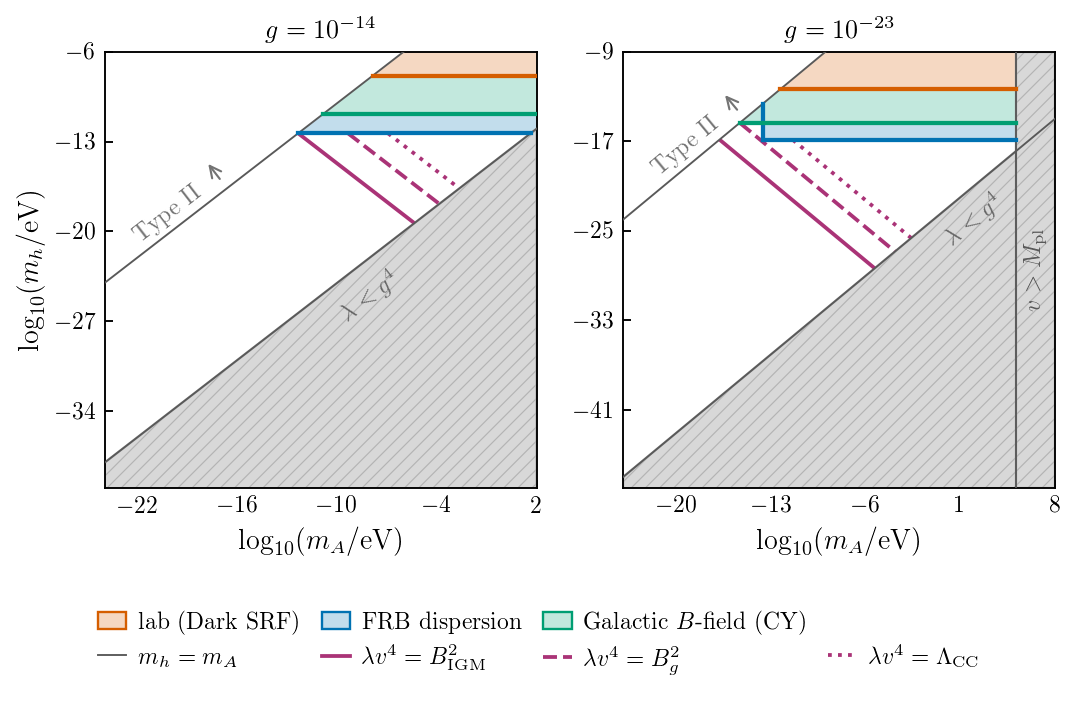}
    \caption{Type I parameter space for fixed $g=10^{-14}$ or $10^{-23}$.  In the top left, the parameter space is no longer Type I parameter space.  In the bottom right (far right) grayed out region, $\lambda \lesssim g^4$ ($v > M_{\rm pl}$) and is theoretically inaccessible.  The blue/green/orange lines indicate the bounds coming from FRBs/galactic magnetic fields/lab experiments.  The $m_h$ dependence of these bounds comes from imposing that symmetry is not restored, $B < B_c$.  Finally the purple lines indicate when the vacuum energy of the higgs field is equal to various other energy densities of interest. Small isolated regions with photon masses up to $10^{-5}\,{\rm eV}$ can be present in the Universe, corresponding to the intersection of the purple solid line and the $\lambda = g^4$ line.
    }
    \label{Fig:Type1}
\end{figure}

All experiments testing the mass of the photon are performed in a background magnetic field. On Earth, there is the $\sim {\rm G}$ terrestrial magnetic field.  In the galaxy there is the $\sim {\rm \mu G}$ galactic magnetic field.  If these magnetic fields exceed the critical value $B_c = \lambda v^2/g$, the symmetry is restored and no constraint can be placed
on the vacuum photon mass.
Even in a magnetically shielded experiment, if the
shielded region is smaller than $1/m_h$, gradient terms hold $v = 0$, and the photon remains massless even though the field has been screened away. These
concerns are what limit tests of the photon mass in the Type I scenario. A summary of the constraints at fixed $g$ is shown in Fig.~\ref{Fig:Type1}.

Before discussing the results in Fig.~\ref{Fig:Type1} in detail, let us summarize the conditions that bound the parameter space. First, the millicharged higgs is extremely light in the Type I parameter space, and $g < 10^{-14}$ is required to satisfy the stellar cooling bounds~\cite{Fung:2023euv}.
Second, we require $\lambda/g^2<1/2$ ($m_A>m_h$), which is what defines a Type I photon mass.
Third, we require $\lambda > g^4$, since even if $\lambda$ were set smaller, RG evolution would quickly drive it back to $\lambda \sim g^4$ (see Sec.~\ref{sec:2typeI} for more details). 
Fourth, we require $B < B_c$; otherwise the symmetry is
restored and no constraint can be placed on the vacuum photon mass. For fixed $g$, these are horizontal lines in the parameter space (fixed $m_h$). Finally, we require $v < M_{\rm pl}$. 

There are several striking features in these plots.  As mentioned before, it is actually possible for the photon to have a mass in vacuum that is as large as $3 \times 10^5 \,{\rm eV}$, as shown in the right hand plot of Fig.~\ref{Fig:Type1}. 
The second feature, as mentioned previously, is that the millicharged higgs has a maximum mass of $m_h \lesssim 10^{-13} \, {\rm eV}$ in the Type I parameter space, stemming from the requirement that $B_{\rm IGM} > B_c$ so that the symmetry is restored everywhere in the Universe. $m_h$ is maximized when $g \simeq 10^{-14}$ and the magnetic field in extra galactic space $B_{\rm IGM} = B_c$.  This situation is shown in the left hand plot of Fig.~\ref{Fig:Type1}.

A last feature worth emphasizing in Fig.~\ref{Fig:Type1} are the three purple lines, which show a comparison between the higgs potential energy and the energy density (pressure) in the intergalactic and galactic magnetic fields, as well as the cosmological constant. In this case, if a symmetry breaking bubble ever nucleates anywhere in the universe, the surrounding magnetized medium cannot exert enough pressure to stop its expansion. Such a possibility might be too interesting, as it can totally destabilize the universe. On the other hand, to the left the purple line, these bubbles can be stabilized by the magnetic pressure after their formation, and leave interesting observational signatures we elaborate on in Sec.~\ref{sec:SCsky}

Finally, it is worth mentioning that most of this region of parameter space is technically un-natural.  A technically natural theory would obey $m_h > g \, {\rm TeV}$, where new physics, e.g. superpartners, are taken to appear at the TeV scale to render the theory natural.  Combined with $B_{\rm IGM} > B_c $ shows that the theory can only be technically natural when $g < 10^{-35}$.

\subsubsection{Type II}\label{sec:4typeII}

\begin{figure}
    \centering
    \includegraphics[width=0.8\linewidth]{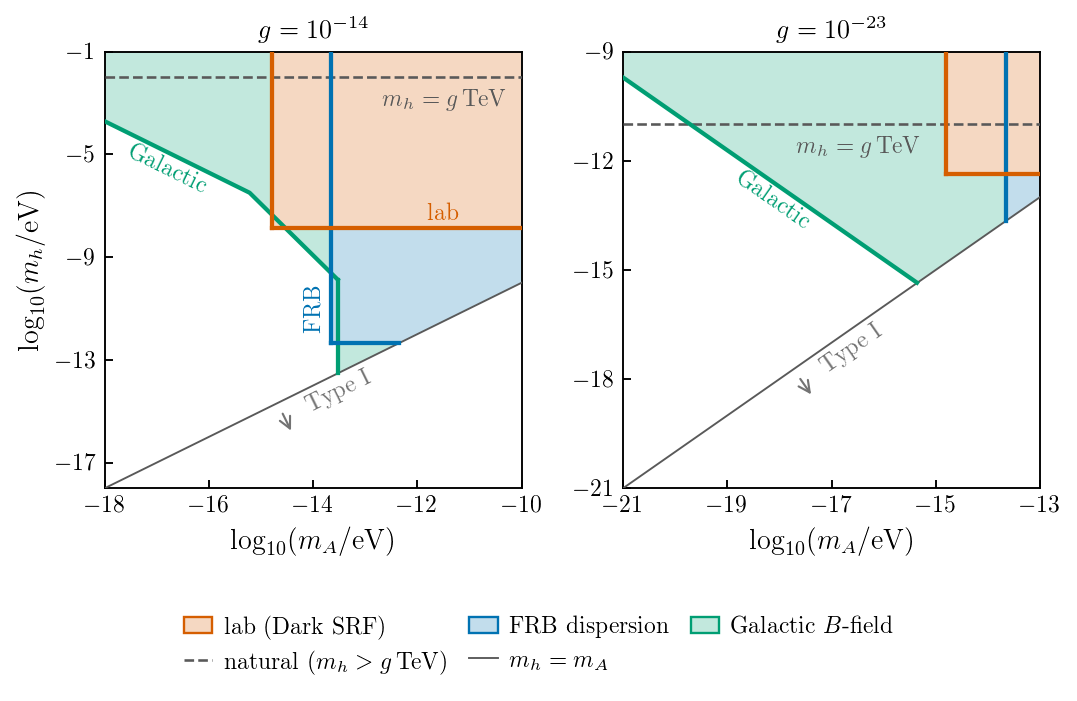}
    \caption{Type II parameter space for fixed $g=10^{-14}$ or $10^{-23}$.  In the bottom right, the region is no longer Type II parameter space.  The blue shaded region is excluded by FRBs.  Below it, intergalactic magnetic fields restore the symmetry and no experimental constraints apply.  
    The green line represents galactic constraints coming from requiring that the dynamo process can produce the galactic magnetic fields, see Sec.~\ref{sec:plasma}.
    The solid orange line represents lab based constraints and below it, Earth's magnetic field restores the symmetry.  Finally, the region above the black dashed line is technically natural.}
    \label{Fig:Type2}
\end{figure}

As in the Type I case, we first summarize when each constraint applies, and then discuss the details of Fig.~\ref{Fig:Type2}. Measurement of Fast Radio Bursts (FRBs) constrain the photon mass through the massive-photon dispersion relation as the signal propagates through the intergalactic medium~\cite{Bonetti:2017pym,Wang:2023fnn}.
These bounds therefore disappear only when the symmetry is fully restored, i.e. when
$B_{\rm IGM} > B_{\rm c2}$, and are shown as the blue
shaded region in Fig.~\ref{Fig:Type2}. Traditional galactic $B$-field bounds, by contrast, vanish as soon as vortices can form. The various vortex formation conditions discussed in Sec.~\ref{sec:plasma}, which survive vortex formation, are shown in green.
Finally, Earth-based constraints are insensitive to vortex formation and vanish only for $B_\oplus > B_{\rm c2}$. These lab constraints are shown in orange.

As discussed already in Sec.~\ref{sec:mag}, there is very limited parameter space lab based experiments can probe without shielding off the terrestrial magnetic field. This is because for a fixed $g$, the symmetry restoration on the Earth can only be avoided if the millicharged higgs is heavy enough. On the other hand, increasing the higgs mass for a fixed $m_A$ and $g$ also increases the superheating field strength $B_{\rm sh}$ and critical current $J_c$, making it harder for vortices to form in the galaxy. In order to probe new parameter space with lab experiments beyond the current limit of $m_h \lesssim 10^{-8} \,{\rm eV}$ and $m_A \lesssim 1.6\times 10^{-15} \,{\rm eV}$ (See Fig.~\ref{Fig:Type2}), one could either improve the precision to get to lower photon masses, or, potentially simpler, perform the experiment in environment with smaller magnetic field to get to lower higgs mass. Whereas shielding a room of $\mathcal{O}(100 \,{\rm m})$ significantly below the terrestrial magnetic field strength might be difficult, it is conceivable that one can design and perform a photon mass measurement in space, where the magnetic field strength can be significantly smaller. We elaborate on these possibilities in Sec.~\ref{sec:lowmag}.
Finally, it is worth mentioning that unlike a Type I photon mass, a Type II photon mass can easily be technically natural. 

\subsection{Novel Experimental Directions}

The theoretical results of Sec.~\ref{sec:mag} also point to new experimental opportunities, both in the laboratory and in astrophysical and cosmological observations. In the laboratory, our analysis of the Type II parameter space motivates creating or traveling to environments with weak ambient magnetic fields and performing experiments there. On cosmological scales, for both the Type I and Type II theories, the photon mass could be as large as $\sim 10^{-14}\,{\rm eV}$ everywhere outside galaxies. For Type I, it is also possible for cosmic voids to contain small isolated regions with photon masses up to $10^{-5}\,{\rm eV}$. These possibilities suggest two observational tests: searches for an anomalous frequency-dependent optical depth in radio observations, and measurements of the dispersion measures of distant fast radio bursts, both their redshift dependence and their correlation with voids along the line of sight.

\subsubsection{Photon mass measurement in a low magnetic field region}\label{sec:lowmag}

In Sec.~\ref{sec:4typeII}, we showed that existing lab experiments on Earth can only probe very limited region of parameter space due to the presence of the terrestrial magnetic field and existing strong constraints from the galactic dynamics (see Fig.~\ref{fig:typeIIenv} and Fig.~\ref{Fig:Type2}). On the other hand, our study also points to a new experimental direction. Besides improving the sensitivity of existing experiments, one can explore
new regions of parameter space in the laboratory by creating an environment,
larger than the inverse higgs mass, that is shielded from the Earth's
magnetic field~\cite{Hook:2017psm}. For a box smaller than the inverse
higgs mass, the higgs field remains in the symmetric phase and no nontrivial profile can form.  For a box larger than the inverse higgs mass, inside the shielded region,
whose boundary we treat as thin and within which the magnetic field is well
below $B_c$, a nontrivial higgs profile develops and the photon becomes
massive. The higgs field profile $\mathcal{H}(x)$ can be computed analytically in 1D in the limit $g B_{\oplus} \gg m_h^2$,
\begin{equation}
    \mathcal{H}(x) = \sqrt{\frac{2}{\lambda}}\frac{m_{h}k}{\sqrt{1+k^2}} \mathrm{sn} \left(\frac{m_h x}{\sqrt{1+k^2}},k\right),
\end{equation}
where $\mathrm{sn} ()$ is the Jacobi elliptic sine function, $x\in[0,L]$, and
$k$ is determined by $m_{h} L = 2\sqrt{1+k^2}\,K(k)$, with $K(k)$
the complete elliptic integral of the first kind. Since
$2\sqrt{1+k^2}\,K(k)$ increases monotonically from $\pi$ at $k=0$, a
solution exists if and only if $m_{h} L > \pi$, confirming the earlier
statement that a non-trivial higgs profile is possible only for
sufficiently large boxes.
For a cubic box of side $L$, the higgs field profile can only be obtained numerically, but the threshold for the existence of such a profile can be found analytically to be $m_{h} L = \sqrt{3}\pi$. 

The limiting factor for these experiments is that the mass of the higgs is bounded from above because
\begin{equation}
    B_c \simeq m_h^2/g \leq B_{\oplus} \quad \rightarrow \quad m_h \leq \left(g  B_{\oplus} \right)^{1/2} = 10^{-8} \,{\rm eV}\left(\frac{g}{10^{-14}} \right)^{1/2}\left(\frac{B_{\oplus} }{{\rm G}}  \right)^{1/2}.
\end{equation}
This means that in order to explore the parameter space, one need to create magnetically shielded regions with field smaller than $B_c$ with sizes of roughly $(100 \,{\rm m})^3$ or larger. Achieving this on Earth requires dramatically increasing our ability to shield a room from Earth's magnetic field. Current $\mu$-metal shield can achieve $\mathcal{O} ({\rm \mu G})$ ambient magnetic field in a box of about 3 meters~\cite{Altarev:2015fra}, and smaller magnetic field down to $\mathcal{O} ({\rm n G})$ can be achieved by utilizing superconducting shield~\cite{cabrera1975use,Fadeev:2020gjk}. A second option, which seems much simpler, is to perform a photon mass measurement away from the Earth. For example, near the Lagrange points, the interplanetary magnetic field (IMF) carried by the solar wind at $1 \,{\rm AU}$, is $\lesssim 10^{-4} {\rm G}$, and this further decreases to $\lesssim 10^{-5} {\rm G}$ around the location of Voyager~\cite{2013LRSP...10....5O,2021ApJ...906..119B}. 

Exploration of Type I parameter space requires much more futuristic technology. Since $m_h\lesssim 10^{-13}\,{\rm eV}$, one would need to hold the field below ${\rm \mu G}$, or even  ${\rm n G}$, over a box of size at least $\sqrt{3} \pi (10^{-13} \,{\rm eV})^{-1} \approx 10^4 \,{\rm km}$, which is certainly beyond reasonable.

\subsubsection{Superconductors in the sky}\label{sec:SCsky}

So far, we have assumed a magnetic field strength in the IGM of $\,{\rm n G}$ to map out qualitatively the parameter space of the model. Observationally, limits on intergalactic magnetic field exists when the correlation length is long~\cite{Pshirkov:2015tua}, or if this field is primordial in nature~\cite{Planck:2015zrl,Durrer:2013pga}. However, if magnetic fields exist on very small scales, then there is very little we know about the strength of the intergalactic field~\cite{Pshirkov:2015tua,Neronov:2010gir}. 

More importantly, these intergalactic, or even galactic magnetic fields can have strong spatial variability. This translates into spatial-temporal variability of the photon mass. In particular, in the case of Type I, there is the possibility of some period of the early universe evolution, or some regions of the universe today, where $U(1)_{\rm EM}$ gets broken, and the photon acquires a large mass, if regions of small magnetic field can exist with sizes larger than $1/m_h$

If such regions exist in the Universe, then there can be two distinct behaviors depending on whether the vacuum energy released in the phase transition ($\lambda v^4$) is larger or smaller than the magnetic field and gas pressure density of the surrounding medium. 
If $\lambda v^4$ is larger (parameter space to the right of the magenta solid line in Fig.~\ref{Fig:Type1}), then the Meissner phase will grow in size.  This region could potentially overtake the whole Universe in the region above the dashed magenta line. Between the dashed and the solid magenta line, this could also happen if significant momentum builds up on the bubble wall before  colliding with a galaxy. Such a runaway event would lead to catastrophic consequences in the parameter space much above the magenta dotted line, where the vacuum energy released could be so large that the inside of the bubble becomes Anti de-Sitter space and inevitably crunches.  
If we were found to be existing in this region of parameter space, it is recommended for future generations to avoid building a magnetically shielded box with size $\sim 10^3 \,{\rm km}$ and strength well below $\,{\rm nG}$.  It would trigger a phase transition and possibly destroy the Universe~\cite{Kobzarev:1974cp,Coleman:1977py}. 

To the left of the magenta solid line in Fig.~\ref{Fig:Type1}, these fluctuations lead to localized regions of Meissner phase. These regions behaves like a superconductor, where photons with frequency lower than $m_A$ will bounce off the superconductor surface. The maximal $m_A \approx 10^{-5} \,{\rm eV} \approx {\rm GHz}$ can be found easily by saturating both $\lambda = g^4$ as well as $\lambda v^4 = B_{\rm IGM}^2$. 
For photons with frequency smaller than $m_A$, scattering off these isolated regions resembles Thomson scattering, with scattering cross section equaling to the geometrical area of these superconducting objects. For photons with frequency higher than $m_A$, these regions can act as a lens, similar to plasma lenses that exist in the Universe. 
Observationally, these regions correlate with locations with the lowest magnetic field, and are most likely to be found in cosmic voids. An excess of frequency dependent screening or plasma lensing that is correlated with voids would be a smoking gun signal of the existence of these rare objects \cite{1980ARA&A..18..537S,Dvorkin:2008tf}. 

\subsubsection{Fast Radio Burst}

Another way to explore the parameter space below the green solid line is by using Fast Radio Bursts (FRBs), neutron star mergers, and other sources of light at cosmological distances. 
Distant sources of light are sensitive to the photon mass through the frequency-dependent delay it imprints on their arrival times, i.~e., the dispersion measure (DM). With a nonzero photon mass, the dispersion measure is
\begin{equation}
    {\rm DM} = \int \frac{{\rm d} z}{(1+z)^2 H(z)} \left(n_e(z) + \frac{m_e}{e^2} m_A^2\right).
\end{equation}
These searches have been done in~\cite{Bonetti:2017pym,Wang:2023fnn}, with different assumptions about the galactic contribution, the host galaxy contribution and their redshift dependence. With a population of sources extending to high redshift, the photon mass can be distinguished from the plasma frequency ($n_e(z)$) through its different redshift dependence, since $n_e(z)$ scales as $ (1+z)^3$.

Our study also points to a qualitatively different feature, coming from medium dependence of the photon mass. For $g B_{\rm IGM} < m_h^2 < g B_g$, the photon is only massive outside the galaxy. In this part of the parameter space, the photon mass is anti-correlated with the location of the galaxies, and new searches based on the same techniques as~\cite{2026ApJ...998..109S} can be designed to explore this part of the parameter space. We leave a detailed study to future work.

\section{Conclusion}\label{sec:con}

In this work, we have revisited the bounds on the photon mass when the mass is the result of an Abelian Higgs mechanism, as opposed to the commonly assumed Proca mass.  The phenomenology varies greatly depending on if the photon mass is Type I ($\lambda/g^2 < 1/2$) or Type II ($\lambda/g^2>1/2$).  Depending on the type of the photon mass, different phases become available with distinct experimental implications.

For Type I photon masses, the most experimentally interesting parameter space is when galactic magnetic fields restore the symmetry and intergalactic fields allow for small regions of symmetry non-restoration to form. In these regions, the photon mass can be as large as $\sim 10^{-5}\,{\rm eV}$. These regions can reflect low frequency light and act as a lens for higher frequency light, giving unique observable signatures that are correlated with cosmic void.  The most theoretically interesting aspect of this parameter space is that a photon mass as large as $3 \times 10^5$ eV is allowed by data, as everywhere in the universe would be in a symmetry restored phase.

For Type II photon masses, theoretically, we found that in spite of the findings in~\cite{Adelberger:2003qx}, the galactic magnetic field still places the strongest constraints on most of the parameter space. These constraints stem from the difficulty to produce large scale coherent magnetic field in the galaxy, instead of the energy density arguments in~\cite{Chibisov:1976mm,10.1143/PTPS.11.1}.
These new understandings also point to new directions for lab experiments to make progress in. 
Experimentally, the most interesting regime is when the Milky Way is in the vortex phase, while the Earth's magnetic field creates symmetry restoration.  In this case, a non-zero photon mass could be probed by performing photon mass measurements either inside large magnetically shielded rooms or in space, where the system can live in the vortex phase and photon mass can be non-zero. 

Finally, contrary to common perception, limits on the photon mass that are most robust against environmental effects actually come from cosmological measurements, especially from voids in our Universe, since these are likely the region with the smallest magnetic field, and as a result, most likely to support a region in the Meissner phase. 


\begin{acknowledgments}

We thank Asher Berlin, Saptarshi Chaudhuri, Harrison Siegel, Kendrick Smith, Kevin Zhou and Muni Zhou for helpful discussions, and Mehrdad Mirbabayi for comments on the draft. JH would like to thank Tingyu Li for sharing enlightening simulation results and for valuable discussions. JH would also like to thank NYU CCPP for hospitality during the inception of this work.  
We thank Claude Code for stylistic suggestions on the figures in this draft, especially for making them color blind friendly. AH is supported by NSF grant PHY-2514660 and the Maryland Center for Fundamental Physics.
Research at Perimeter Institute is supported in part
by the Government of Canada through the Department
of Innovation, Science and Economic Development and
by the Province of Ontario through the Ministry of Colleges and Universities.
\end{acknowledgments}

\appendix

\section{Magnetostatics at $m_{A} \neq 0$}\label{Sec:static}

In this appendix, we review some of the key concepts in the Abelian Higgs theory of photon mass. We will first review the critical fields and currents in the absence of external sources of current, with a focus on the Type II case. We will then discuss a few toy examples of magnetic fields and vector potentials around external current sources, designed with the galactic magnetic fields in mind.

\subsection{Critical fields and critical currents}

Three critical quantities play a central role in this paper: $B_{\rm c2}$ ($B_{\rm c}$), $B_{\rm sh}$ and $J_c$. When $B > B_{\rm c2} = \lambda v^2/g$, symmetry is restored so that the higgs vev goes to zero and the photon mass vanishes everywhere in the magnetic field.
The simple way to understand this limit involves going back to the Landau level of charged scalars. In a background magnetic field, the mass of the higgs boson in the symmetry restored phase is
\begin{equation}
    m_h^2/2 = g B -\lambda v^2.
\end{equation}
If $gB < \lambda v^2$, the tachyonic mass results in a vev for the higgs.
When $g B >\lambda v^2$, the positive energy of the lowest Landau level overcomes the negative higgs mass squared and the symmetry restored point becomes stable.  

The more important critical quantities for understanding the critical behaviors of this model in our galaxy are the superheating field strength $B_{\rm sh} = \lambda^{1/2} v^2$, and critical current density $J_c = m_A m_h v \simeq m_A B_{\rm sh}$. 
Above these critical values, the system is classically unstable to the formation of strings due to instabilities identified by Galaiko and Feynman, respectively~\cite{Galaiko1966FormationOV,feynman1955chapter,PhysRevLett.19.822} (see also~\cite{East:2022rsi,Fedderke:2025sic} for more detail). 
The critical field $B_{\rm sh}$ and critical current density $J_c$ are, up to $\mathcal{O}(1)$ geometrical coefficients, reached almost always simultaneously in condensed matter systems. This is due to the skin effect, which dictates that the current flows within a distance $R \simeq 1/m_A$ from the boundary of the superconductor, and therefore, $A \simeq B/m_A$ and the London current $ J_L = m_A^2 A \simeq m_A B$. In the following, we will study a few examples of (external) current densities that are motivated by the realistic situation in a galaxy, and show that the relation of $A \simeq B/m_A$ is quite generic if the source of the magnetic field is external to the region with $m_A \neq 0$. We will also consider examples where the relation of $A \simeq B/m_A$ gets violated, and the critical current density can play a more significant role. 

\subsection{Magnetic field and Vector potential around a current source} \label{Sec:outside}

To start with, let us consider an infinite current carrying wire with current $I$ in the $z$-direction, in a medium with $m_A \neq 0$. The resulting magnetic field and vector potential in cylindrical coordinate $(r,\theta,z)$ can be found to be:
\begin{equation}
    \vec{B}(r) = \frac{m_A I}{2\pi} K_1(m_A r)\hat{\theta}, \quad  \vec{A}(r) = \frac{I}{2\pi} K_0(m_A r)\hat{z}.
\end{equation}
Close to the current wire $m_A r \ll 1$, we have the amplitude $B \approx A/r$, just like in the case of $m_A=0$. However, since the modified Bessel function of the second kind $K_{\nu}(r) \sim \exp[-m_A r]$ when $m_A r \gg 1$, both magnetic field and vector potential are exponentially suppressed far away, and $B \approx m_A A$ far away from the wire. 

Similarly, for an infinite current sheet with surface current density $K \hat y$, we find that 
\begin{equation}
    \vec{B} = \frac{K}{2} {\rm sign (z)} e^{-m_A |z| } \hat{x}, \quad  \vec{A} = \frac{K e^{-m_A |z| }}{2 m_{A}} \hat{y}.
\end{equation}
Again, we find that both magnetic field and vector potential are exponentially suppressed far away, and $B \approx m_A A$ far away from the current source. 

The combination of these two examples suggests that when the source of the current is outside the region where the field is present, then we expect $A \lesssim B/m_A$ far away from the current source, instead of $A \sim B R$, where $R$ represents the distance from the source of current. 

The physical implication of these two simple examples is that, in order for the galaxy to maintain its large scale coherent magnetic field in the Meissner phase of a theory with $m_A R_{\rm galaxy} \gg 1$, the current that supports this large scale magnetic field must be flowing inside the bulk of the galactic medium itself.

\subsection{Current-Field relation in a plasma} \label{Sec:inside}

In light of the findings in the previous subsection, we consider a third simple example with a volume filling external current density
\begin{equation}
    \vec{j}_{\rm ext}(x) = j_0 \sin[x/ R] \hat{y},
\end{equation}
with $m_A R\gg 1$, then the resulting vector potential and magnetic field can be found as
\begin{equation}
    \vec{B}(x) = \frac{j_0 R}{1+ m_A^2 R^2}\cos[x/ R] \hat{z}, \quad  \vec{A}(x) =\frac{j_0 R^2}{1+ m_A^2 R^2}\sin[x/ R] \hat{y}.
\end{equation}
In this case, the magnetic field is roughly a constant for $x \ll R$, over a range that could be much larger than $1/m_A$.  As is clear from this solution, one can maintain the relationship $A \sim B x$ over a range $\gg 1/m_A$.  The price is simply that the source itself must extend into the region under consideration.

Evidently, when $m_A R \gg1$, the London current density and the magnetic field are no longer related by $ J_L \simeq m_A B$. In fact, for $x/R \sim 1$, $J_L \rightarrow j_0$ and $ B \rightarrow 0 $ in the limit of $m_A R\rightarrow \infty$. This is a manifestation of the Meissner effect, where the London current density almost cancels the large scale external current density, such that the magnetic field inside the superconductor can remain close to zero. Quantitatively, we have 
\begin{eqnarray} \label{Eq:J}
    \vec J_L &=& - m_A^2 \vec{A}  = - \frac{m_A^2 R^2}{1+m_A^2 R^2}\vec{j}_{\rm ext} \, \\ 
    \vec j_{\rm tot} &=& \vec J_L + \vec{j}_{\rm ext} =  \frac{1}{1+m_A^2 R^2}\vec{j}_{\rm ext}.
\end{eqnarray}
Since the magnetic field is generated by $j_{\rm tot}$, it is significantly suppressed in the limit of $m_A R \rightarrow \infty$. In such a system, it is much easier to reach the critical London current $J_c$ than the critical magnetic field $B_{\rm c2}$. In fact, we find that at string formation,
\begin{equation} \label{Eq:R_string}
     B_{J_c} \approx \frac{B_{\rm sh}}{m_A R} \, .
\end{equation}
This condition was used in~\cite{Adelberger:2003qx} as the critical field for producing a network of vortices in the galaxy, and $B_{g} >  B_{J_c}$ for $R = R_{\rm galaxy}$ is chosen as the sufficient condition for evading the Chibisov-Yamaguchi bound~\cite{Chibisov:1976mm,10.1143/PTPS.11.1}.

This is actually not true, since it hides the main difficulty in this scenario. In order to produce a large scale magnetic field of $B_{g} \approx {\rm \mu G}$ for $R = R_{\rm galaxy} \approx {\rm kpc}$, the current density required is enormous. For example, when $m_A = 10^{-15} \,{\rm eV}$, we have
\begin{equation}
    j_{\rm ext} \approx m_A^2 B_g R_{\rm galaxy} \simeq 400 /{\rm cm^3},
\end{equation}
which cannot be attained in a galaxy with charge particle density of $\lesssim {\rm cm^{-3} }$. This exemplifies the importance of understanding the dynamics in the magnetized plasma of a galaxy.

\section{Magneto-hydrodynamics for Particle Physicists - $m_A = 0$} \label{Sec:review}

In this appendix, we provide a very brief review of $m_A =0$ magneto-hydrodynamics for particle physicists (see also~\cite{Rincon:2019coh} and references within).  The main purpose is to introduce the induction equation, the Reynolds numbers $R_m$ and $R_e$ that determine when a dynamo can operate, and the Kolmogorov scalings that set the smallest length scale to the turbulence, providing the context for the $m_A\neq 0$ generalization in the main text.

\subsection{The induction equation and the magnetic Reynolds Number ($R_m$)} 

Our starting equations are Maxwell's equation and Ohm's law
\bea
\nabla\times\vec E &=& - \frac{\partial\vec B}{\partial t} \\
\nabla\times\vec B &=& \vec j +  \frac{\partial E}{\partial t} \approx \vec j \label{eq:B2}\\
\vec j &=& \sigma\!\left(\vec E + \,\vec u\times\vec B\right).
\eea
with $\sigma$ the conductivity and $\vec u$ the fluid velocity. 
It can be shown that we can drop the displacement current in Eq.~\ref{eq:B2} in the non-relativistic limit ($u\ll 1$).
Combining these equations, we find the induction equation
\begin{equation} \label{Eq:MHD}
\frac{\partial\vec B}{\partial t} = \nabla\times(\vec u\times\vec B) + \eta\,\nabla^2\vec B
\end{equation}
with $ \eta = 1/\sigma$, which is the central equation of magneto-hydrodynamics.

The two terms on the right-hand side compete. Acting alone, the second (diffusion) term causes magnetic field to decay away,
\begin{equation}
\vec B(t) = e^{- \eta k^2 t} e^{i k x} \vec B_0.
\end{equation}
The first term, on the other hand, locks the magnetic field into the fluid. For a surface whose boundary is advected with the fluid,
\bea \label{Eq:Perfect_Induction}
\frac{d \Phi_B}{dt} = \int \frac{\partial \vec B}{\partial t} \cdot dA + \oint_C B \cdot (\vec u \times d \vec l) = \int \nabla\times(\vec u\times\vec B) \cdot dA - \int_c (\vec u\times\vec B) \cdot d \vec l = 0,
\eea
where the first term captures the change of magnetic field through the surface and the second the motion of the boundary.
If only the first term is present in Eq.~\ref{Eq:MHD}, by Stokes' theorem, $\frac{d \Phi_B}{dt} = 0$. This is flux freezing (Alfv\'{e}n's theorem), flux lines move with the fluid, so contraction amplifies the field while expansion dilutes it.

There is a second way to write Eq.~\ref{Eq:MHD} that is useful.  By decomposing the $\nabla\times(\vec u\times\vec B)$ term into 3 pieces and defining $D/Dt = \partial/\partial t + \vec u \cdot \nabla$, we obtain
\begin{equation}\label{eq:MHD2}
\frac{D\vec B}{Dt} = (\vec B\cdot\nabla)\vec u - \vec B\,(\nabla\cdot\vec u)+ \eta\,\nabla^2\vec B.
\end{equation}
Physically, the term $(\vec B\cdot\nabla)\vec u$ is the stretching (dynamo) term partially responsible for the growth of the magnetic field, while $-\vec B(\nabla\cdot\vec u)$ is the compression term. This rewriting makes it clearer that magnetic field growth requires particular properties of the fluid velocity profile. For a few examples of dynamo processes, see~\cite{Rincon:2019coh} and references within.

Which of the two terms dominate at a length scale $L$ depends on the first important dimensionless number, the Magnetic Reynolds number $R_m$, defined as the ratio of these two terms
\bea\label{eq:Rm}
R_m = \frac{ u/L}{\eta/L^2} = \frac{u L}{\eta} .
\eea
If $R_m \gg1 $, flux is frozen and the field is stretched and compressed with the fluid. Otherwise, if $R_m \ll 1$, the field simply decays.

\subsection{Turbulence, the viscous scale, and the Reynolds number ($ Re$)}

Moving on to the dimensionless number that characterize the fluid itself, the (kinetic) Reynolds number is
\bea
Re = \frac{\rho u L}{\mu} = \frac{u L}{\nu} \, ,
\eea
with $\nu$ the kinematic viscosity parameter. The Reynolds number measures the importance of inertia as compared to viscous damping. $Re \gg 1$ is the turbulent regime, which is the case of interest since Laminar flow is generally too smooth to provide the velocity profile that can drive a dynamo. A third dimensionless number that is useful for describing dynamo is the magnetic Prandtl number, $P_m \equiv R_m/Re = \nu/\eta$, an intrinsic property of the fluid. The magnetic Prandtl number compares the two diffusivities, and $P_m \gg 1$ would ensure that the system evolution is dictated by the kinematics of the velocity field, which is indeed the case for the plasma in the galaxy.

The small-scale dynamo operates in the background of Kolmogorov turbulence. Kinetic energy injected at the outer scale $\ell_0\simeq \,{\rm kpc}$, and subsequently cascades to smaller scales at a constant rate $\epsilon$ per unit mass. On dimensional grounds, the velocity on scale $\ell$ is
\begin{equation}
u (\ell)\sim (\epsilon \ell)^{1/3} = u_0 (\ell/\ell_0)^{1/3},  
\end{equation}
which is the scaling used in Sec.~\ref{sec:plasma}. The cascade terminates at the viscous (Kolmogorov) scale $\ell_\nu$, where the Reynolds number drops to unity ($ Re (\ell_\nu) = u(\ell_\nu) \ell_{\nu}/\nu =1$). Since the viscosity of a fluid is the thermal velocity $\approx u_0$ times the mean free path $\ell_p$, which is roughly $10^{15} \,{\rm cm}$ for neutral hydrogen inside the galaxy, we find that $\ell_{\nu} \approx \ell_0^{1/4} \ell_p^{3/4} = 0.01\,{\rm pc}$ in the warm interstellar medium, the smallest scale at which the turbulent eddies can be generated, and hence, the dynamo can operate.

\subsection{Small Scale Dynamo}

The small-scale dynamo is commonly understood through the geometric Stretch-Twist-Fold mechanism. Take a closed loop of magnetic field and stretch it to twice its length, due to flux freezing, the cross section of the flux tube halves and the magnetic field strength doubles to conserve flux. The magnetic loop is then twisted into a figure eight, and folds back into a circle of the original size. In this process, one ends up at the original shape, with the same original area, but with twice the magnetic field. Repeating this process exponentially enhances the magnetic field. Turbulence is excellent for implementing this Stretch-Twist-Fold mechanism since it is chaotic.

As the details of the small scale dynamo are too complicated to be discussed here, we end by simply quoting a set of sufficient (though not strictly necessary) requirements for it to operate on a scale $\ell$
\bea \label{Eq:SSdynamo}
R_m (\ell) \gtrsim 1 \qquad P_m \gtrsim 1 \qquad Re (\ell)\gtrsim 1\, .
\eea
The first two conditions ensure the magnetic flux is frozen into the fluid, while the last ensures that the fluid flow at the scale is turbulent. On every scale satisfying these conditions, magnetic field can grow until it reaches equipartition~\cite{Kulsrud:1992rk} with the turbulent kinetic energy, $\beta B(L)^2 = \rho u(L)^2$, with typical $\beta \sim 10-100$. With a non-zero photon mass, the condition $R_m (\ell) \gtrsim 1$ is modified in the Meissner phase, while the other two conditions remain the same. In a warm interstellar medium, they are both generally satisfied down to the scale $\ell_\nu \simeq 0.01\,{\rm pc}$, which we used in the main text.

\bibliography{reference}

\end{document}